\documentclass[%
 reprint,
 amsmath,amssymb,
 aps,
 prl,
]{revtex4-2}

\usepackage{graphicx}
\usepackage{dcolumn}
\usepackage{bm}
\graphicspath{ {./Figures/} }

\begin{document}
\title{
Active contacts create controllable friction 
} 

\author{Rohan Shah}
 \author{Nick Gravish}%
 \email{ngravish@ucsd.edu}
\affiliation{%
 Mechanical \& Aerospace Engineering Department, \\ University of California, San Diego.
}%

\date{\today}

\begin{abstract} 
Sliding friction between two dry surfaces is reasonably described by the speed-independent Amonton-Coulomb friction force law. 
However, there are many situations where the frictional contact points between two surfaces are ``active'' and may not all be moving at the same relative speed. 
In this work we study the sliding friction properties of a system with multiple active contacts each with independent and controllable speed. 
We demonstrate that multiple active contacts can produce controllable speed-dependent sliding friction forces, despite each individual contact exhibiting a speed-independent friction. 
We study in experiment a rotating carousel with ten speed-controlled wheels in frictional contact with the ground. 
We first vary the contact speeds and demonstrate that the equilibrium system speed is the median of the active contact speeds. 
Next we directly measure the ground reaction forces and observe how the contact speeds can control the force-speed curve of the system. 
In the final experiments we demonstrate how control of the force-speed curve can create sliding friction with a controllable effective viscosity and controllable sliding friction coefficient.
Surprisingly, we are able to demonstrate that frictional contacts can create near frictionless sliding with appropriate force-speed control.
By revealing how active contacts can shape the force-speed behavior of dry sliding friction systems we can better understand animal and robot locomotion, and furthermore open up opportunities for new engineered surfaces to control sliding friction. 
\end{abstract}

\maketitle

\section{Introduction}

Sliding friction is one of the most important forces in the natural world and yet its physics remain elusive.
The simplest and most widely employed model of dry sliding friction (attributed to both Amonton and Coulomb \cite{Popova2015-gf}) states that the forces between two bodies are proportional to the normal force between them and independent of sliding speed. 
Many experiments from the nano- to macro-scales have highlighted counter-exmaples in which speed-dependent friction arises in dry friction: stemming from phenomena such as plastic material deformation \cite{Tambe2005-sm,Ward2015-sd}, thermal fluctuations in atomic-potentials \cite{Muser2011-gj}, alignment and compatibility of mesoscale textured surfaces \cite{Vanossi2020-tt, Menga2021-ri, Li2011-so, Menga2023-fl}, dynamics of filamentous and fibrillar surfaces \cite{Ward2015-sd, Gravish2010-bz}, macroscopic surface vibrations \cite{Mao2017-eg, Benad2018-rx,Gutowski2012-zs}, or even in rigid bodies sliding with combined translational and rotational motion \cite{Farkas2003-hc}. 
However, in all these examples the frictional dynamics at the contacts are coupled to the overall sliding speed. 

Here, we are interested in the frictional properties of surfaces whose contact points have independently controllable speeds, which can be considered as ``active contacts'' \cite{Chong2023-gy}. 
A fundamental motivating observation for studying active contact friction stems from recent experiments of a walking robot \cite{Chong2023-la}.
When multiple legs of the robot slipped against the ground a speed-dependent traction force was observed, despite the individual contacts having speed independent Amonton-Coulomb friction.
Active contacts are relevant in a number of physical and biological phenomena such as ground-based locomotion \cite{Ozkan-Aydin2020-wc,Chong2022-kz,Chong2023-gy,Wu2024-ol,Zhao2020-tm, Hu2009-kq} or manipulation of objects with multiple contact points \cite{Vose2012-wl,Cutkosky1986-vo}.  
Thus, better understanding of the frictional properties of active contact systems can help shed light on the novel physics of locomotion \cite{Aguilar2016-nx}.
However, the underlying physics and control of active contact frictional systems is still relatively unknown. 
A fundamental challenge of gaining insight into active contact sliding friction is the inherent nonlinear nature of individual sliding frictional forces \cite{Urbakh2004-gs, Dillavou2018-zx, Ruina1983-ye}, and collections of sliding elements coupled together \cite{Carlson1991-jf, Deng2019-ny, Rubinstein2004-aj}. 
Despite the simple Amonton-Coulomb law being one of the earliest forces introduced in an undergraduate physics class, the dynamics of systems with sliding friction can be extremely complex to understand and model. 

\begin{figure}[b]
    \centering
    \includegraphics[width=.8\linewidth]{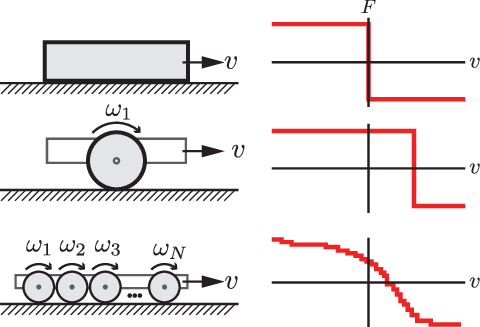}
    \caption{
    Conceptual examples of sliding systems and their force-velocity curves for uniform contact (top row), a single active contact (middle row), and multiple active contacts (bottom row). 
    When each system is moved with net speed $v$ they experience a frictional resisting force $F$ shown in the right column accordingly. 
    }
    \label{fig:motivation}
\end{figure}

\begin{figure*}[t!]
    \centering
    \includegraphics[width = \linewidth]{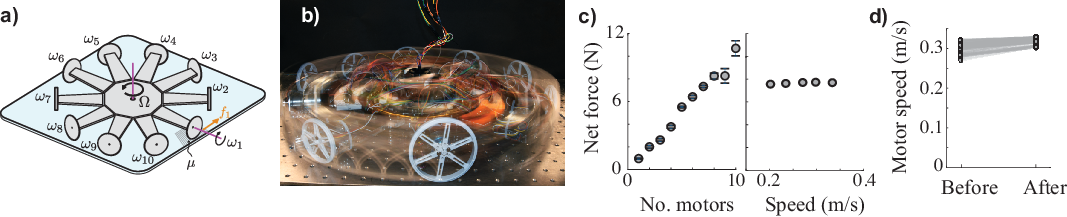}
    \caption{
    System design and friction properties. 
    a) Schematic of multi-contact friction rotational ``carousel''. 
    An array of ten motors drive individual wheels (radius $r_i$) at rotational speed $\omega_i$, measured by encoders on each motor. 
    The entire carousel is free to rotate and a central encoder measures the system rotation rate, $\Omega$.
    c) (Left) The total friction force is linearly dependent on the number of motors that are in contact with the ground. 
    (Right) The total friction force of ten motors with the same contact speed ($r_i \omega_i = r_j \omega_j$) is independent of the motor contact speed.
    d) The speed of individual motors is not affected by slip at the wheel-ground interface.
    }
    \label{fig:setup}
\end{figure*}

In this work we will experimentally demonstrate that systems with multiple active contacts exhibit novel sliding friction phenomena and present opportunities for the modulation and control of sliding friction between two surfaces. 
Our hypotheses are motivated by an initial thought experiment comparing single and multi-contact systems with simple speed-independent friction of each contact. 
Following this example, we introduce an active contact experiment, a frictional carousel, that can individually control the contact speed of ten points.  
Lastly, by varying the individual active contact speeds in experiment we demonstrate that active contacts can shape the overall force-speed dynamics between two dry-friction surfaces.

\textit{Motivating example--}
To motivate the novel physics of active contact friction we envision a representative comparison of three sliding systems (Fig.~\ref{fig:motivation}): 1) a sliding block, 2) a sliding wheel with rotational speed $\omega_1$, radius $R$, and thus constant tangential speed $v = \omega_1 R$, and 3) a system of $N$ wheels each with radius $R$ and independent rotational speeds, $\omega_i$, thus each having tangential speeds $v_i = \omega_i R$. 
We are interested in the relationship between the sliding speed of the overall system, $V$, and the net frictional contact force exerted on the system, $F$, through ground contacts.
All three examples are the same weight and the sliding friction of all individual contacts are governed by the basic Amonton-Coulomb friction law.

The sliding friction force on a rigid block is $F = \mu m g \frac{v}{|v|}$ which is the basic Amonton-Coulomb model of friction (red curve top row, Fig.~\ref{fig:motivation}).
As long as the direction of motion is the same, the sliding friction force is speed independent. 
By adding a single active contact, such as a constant speed wheel, the force-speed curve becomes shifted to the left or right with a force of $F = \mu mg \frac{v_i - V}{|v_i - V|}$.
However, with just one active contact the force-speed relationship is still a ``step-function'' with speed-independent friction.
However, the net force acting on a system with multiple active contacts is governed by the summation of each individual speed-independent contact, 
\begin{equation}
    F(V) = \frac{1}{N}\mu mg \sum_{i = 1}^{N} \frac{v_i - V}{|v_i - V|}
    \label{eq:friction}
\end{equation}
Here we find a peculiar phenomena, by combing many active contacts that move at different speeds the net force-speed curve of the system is no longer restricted to be a step-function but instead can exhibit complex (and controllable) force-speed behavior (Fig.~\ref{fig:motivation}, bottom row). 

To intuitively understand the force-speed relationship in the case of multiple active contacts we imagine that the tangential contact speeds are ordered from slowest to fastest ($v_i < v_{i+1}$). 
At system speed $V < v_1$ all the wheels are actively pulling forward and the net force is $F = \mu m g$, however when $V$ exceeds the speed of the slowest wheel the sign of that friction force changes and now $N-1$ wheels are pulling forward, and one wheel is pulling back resulting in $F = (N - 2) \mu m g$. 
As the system speed is further increased the process will continue and produce a stair-stepped force-speed relationship until ultimately the sliding speed is faster than fastest active contact, $ v_N < V$, and all contacts generate a negative force ($F = - N \mu m g$).
This thought experiment introduces the fundamental observation that multiple active contacts can create controllable force-speed curves.

\begin{figure*}[t]
    \centering
    \includegraphics[width = 1\linewidth]{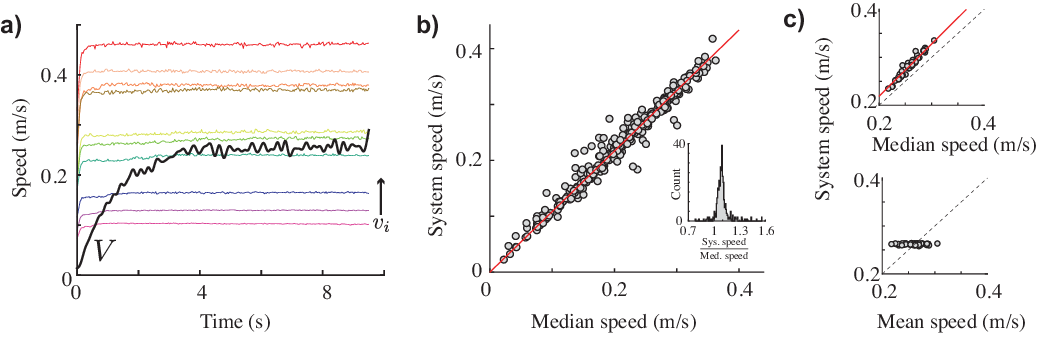}
    \caption{
    Equilibrium behavior of multi-contact frictional systems. 
    a) Speed versus time of the individual active contacts ($v_i$, colored lines) and the system ($V$, black line). 
    After a transient startup period the contacts and system settle into a steady-state. 
    b) The system tangential speed is linearly related to the median speed of the ten wheels over 660 experiments. 
    The red curve is fit line, $y = mx$, with slope $m = 1.065\pm0.01$. 
    The inset shows a histogram of the ratio between system speed and median speed. 
    c) In additional experiments we held the mean of the active contacts constant while varying the median (left plot), or we held the median constant while varying the mean (right plot). 
    These experiments clearly show the system speed is unaffected by the mean wheel speed and only dependent on the median. 
    }
    \label{fig:equilib}
\end{figure*}

\textit{Experiment design and characterization--}
To study the physics of active contact systems we developed a rotational ``carousel''.
The carousel consisted of ten wheels arranged around a central axis, each wheel's rotation speed was generated by an individual DC motor (Fig.~\ref{fig:setup}a,b).
By utilizing a rotational system we are able to study the steady-state motion of the carousel indefinitely, without the space limitations that would be present if studying a similar linear sliding system. 
The rotational carousel was fabricated from laser cut acrylic. 
The central plate of the carousel was mounted on a rotational turntable ``lazy susan'' bearing which provided low rotational resistance while supporting the motor electronics. 
Each ``spoke'' of the carousel was attached via a closet hinge to the central plate to ensure that an equal normal force was distributed across all ten wheels in contact with the ground.
On each of the ten spokes we mounted an individual wheel of constant radius that was independently controlled by an individual DC motor and motor driver. 

The wheels were made of plastic and the ground was acrylic. 
The DC motors (Pololu, \#4866) used a 75:1 gear reduction between the motor and the output wheel.  
The motor speeds are determined by the motor voltages which are controlled through two Teensy microcontrollers which generated pulse-width-modulation (PWM) control signals for five motors each.
Each motor had a magnetic encoder attached to its shaft with a resolution of 48 counts per revolution.
The encoder resolution is multiplied by the gear ratio (75:1) when considered at the wheel contact and thus yielded an equivalent linear resolution of $78$~$\mu$m, however we estimate that through backlash in the gear box the actual resolution is approximately $\approx$800~$\mu$m
The central plate of the carousel used a capacitive rotational encoder (AMT CUI AMT103-V) with 8192 counts per revolution. 
All the encoders and the motor control were sampled at 50~Hz and the data was logged by the Teensy microcontroller over serial communication to computer. 
All electrical signals to the rotational carousel passed through a 12-wire electrical slip ring (DigiKey 1528-1176-ND) which allowed for continuous rotation without wire tangle.

In all experiments we measured the rotational rates of the individual wheels and the overall system. 
To enable comparison between the active contact speed of the wheels and the motion of the carousel we adopt the convention of transforming all rotational rates to their equivalent linear speeds. 
The linear speed of the active contacts are determined by $v_i = R \omega_i$, where $R = 4.5$~cm is the wheel radius and $\omega_i$ the rotational speed of the $i^\text{th}$ wheel.
The system's equivalent linear speed is determined by the carousel rotation rate and the center-to-contact length ($L = 26.6$~cm) given by $V = L \Omega$.

\begin{figure*}[t]
    \centering
    \includegraphics[width=1\linewidth]{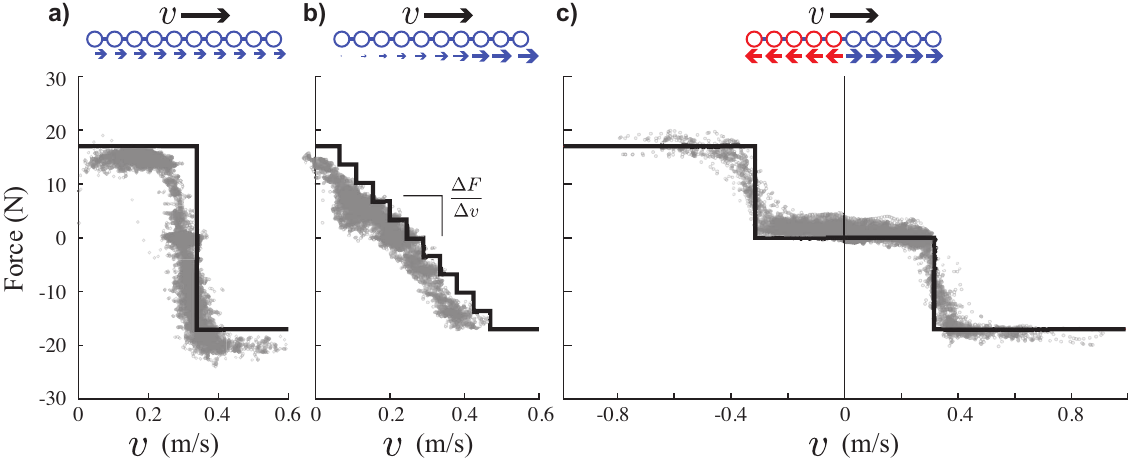}
    \caption{Ground reaction force measurements from active contacts with programmed force-speed curves (black lines).
    In the first experiment all speeds are equal producing a net dry friction force-speed curve. 
    In the next two experiments we commanded different effective viscosity force-speed curves. 
    The last experiment set half the contacts to $-v_0$ and half the contacts to $+v_0$ creating a force-speed regime with near zero force.
    }
    \label{fig:force_vel}
\end{figure*}

To characterize the frictional properties of the carousel we measured the friction force from the wheels when pushing against a stationary load cell. 
The friction force from each wheel is the same and the total friction force is linearly dependent on the number of motors that are spinning in contact with the ground (Fig.~\ref{fig:setup}c).
Furthermore, the friction force is independent of the speed of the wheel as expected for dry Amonton-Coulomb friction (Fig.~\ref{fig:setup}c).

The motors were chosen with a large gear ratio which ensured that the force from the wheel contact friction slipping on the ground had only a small influence on the steady-state speed of each motor. 
We measured this by holding the carousel stationary and then releasing it and measuring the change in individual motor speeds, $\omega_i$.
We observed that when the system was released from being held at rest the speed of each motor increased by 4.6$\pm$2.0\%. 
Thus, our system is capable of generating relatively steady-state contact speeds between multiple points of contact that each behave as ideal, speed independent, dry Amonton-Coulomb friction contacts. 

\textit{Equilibrium properties of multi-contact friction--}
We first sought to determine how the steady-state speed of an active multi-contact system is related to the statistics of the individual contact-speeds. 
In 660 experiments we studied the equilibrium speed of the rotational carousel with randomly chosen wheel speeds. 
In each experiment we selected at random ten speeds for the active contacts ($v_i$) and observed the time evolution of the system speed ($V$) when started from rest.
Figure.~\ref{fig:equilib}a demonstrates the time dependent speeds of the system (black curve) and active contacts (color curves) in a single experiment.
The active contacts reach their steady-state speeds quickly ($\approx$~30~ms) and remain constant with small fluctuations while the system takes a much longer time to approach steady-state.

The equilibrium speed of an active contact system occurs when the friction forces from all contacts sum to zero. 
Since the sliding force from each contact is not speed dependent, the equilirium condition for an active contact system is simply determined by the median of the set of active contact speeds (a similar argument was presented in \cite{Wu2024-ol}).
We find in experiment very good agreement with this prediction: whe system speed $V$ varied linearly with the median speed of the ten motors across a broad range of experiments and motor speeds (Fig.~\ref{fig:motivation}b).
To further illustrate that it is the median value of the contact speeds that sets the steady-state speed we performed experiments in which: 1) the mean of the active contact speeds was held constant while the median of the speeds was varied, or 2) the median of the active contact speeds was held constant and the mean was varied. 
In these subsequent experiments we again observed that the system speed followed the median speed of the active contact speeds (Fig.~\ref{fig:motivation}c, left) even when the mean of the wheel speeds were varied. 
However, when the mean of the wheel speeds was changed while holding the median constant the system speed did not change 
 (Fig.~\ref{fig:motivation}c, right).

These first experiments highlight some of the novel features of sliding friction with active contacts. 
For example, since the median of a set is unchanged when multiplied by a constant value this suggests that the equilibrium speed of active contact systems are unaffected by changing the coefficient of friction in Eq.~\ref{eq:friction}.
This has been observed experimentally recently in walking experiments with robots on different textured substrates \cite{Chong2023-la}.
Furthermore, the median of a set is remarkably robust to perturbations and is unaffected by large outliers.
We tested this by allowing the fastest active contacts to vary sinusoidally in time while keeping the median of the active contacts constant. 
The system speed was unaffected by these outlier contact speeds and held its steady state at the median value of the active contacts.

\textit{Measurement of force-speed curves--}
In our motivating example (Fig.~\ref{fig:motivation}) we hypothesized that control of active contact speeds can generate a controllable friction force-speed relationship. 
We next set to directly measure the force-speed relationship in our active contact system. 
A torque sensor (Futek, TFF400) was mounted on an optical table and the acrylic ``ground'' and carousel were mounted atop the sensor. 
By manually rotating the system quasi-statically while the wheels are rotating we are able to directly measure the force-speed curve of our active contacts system.  

We first set all active contacts to have identical speed ($v_i = v$) and measured the force-speed relationship. 
We predict that the force-speed curve for this experiment will resemble that of a single active contact (Fig.~\ref{fig:motivation}). 
In experiment we observed this force-speed step function (Fig.~\ref{fig:force_vel}a) which matched the commanded force-speed profile well (shown in red, Fig.~\ref{fig:force_vel}a). 
We observed slight deviation from the vertical transition between positive and negative force which we attribute to backlash within the motor drive system and the carousel ``arms'' as the friction force reverses direction. 

We next set the active contacts to have varied and uniformly spaced speeds to generate controlled linear force-speed dependence, in effect controlling the interaction ``viscosity'' with the ground. 
We characterize the active contact ``viscosity'' through the linear slope of the force-speed curve, $\frac{\Delta F}{\Delta v}$.
We observed good agreement between the commanded ``viscosity'' and that observed from the force-speed measurements. 
This experiment highlights the first example of how an active multi-contact frictional system can generate and control the force-speed curve between two surfaces.

Lastly, since the equilibrium speed of an active multi-contact system is determined by the median of the contact speeds there is a peculiar phenomena that arises dependent on whether there are an even or odd number of frictional contacts. 
A set with an odd number of elements always has a well defined median whereas this is not always the case for an even number set. 
In sets with even numbers of members the median may not be unique and can instead span a range within the set. 
For example, consider a system with four contact speeds $v = \left[-2, -1, 1, 2 \right]$.
There is no unique median speed of this set and thus the equilibrium system speed ($F = 0$) can be anywhere in the range $V = \left(-1, +1\right)$. 
For a multi-contact frictional system when the median is not defined this effectively means that when the system slides at speeds within the median range it should experience zero frictional force from the ground. 

We tested this hypothesis by setting five active contacts to $v_{min} = -0.35$~m/s and an opposing five active contacts to $v_{max} = +0.35$~m/s.
The force-speed curve from this experiment yielded a stepped pattern consistent with the model prediction (data in blue, model in red Fig.~\ref{fig:force_vel}c).
Remarkably when the system is sliding within the median range of the active contacts it experiences extremely low ground reaction forces despite all the wheels maintaining contact and friction with the ground. 
Thus we hypothesize that controlling active contact speeds can enable control of the overall sliding friction coefficient of the system.
In the next section we demonstrate control of the ``viscosity'' and friction coefficient through active contact speed control.

\begin{figure}
    \centering
    \includegraphics[width=.8\linewidth]{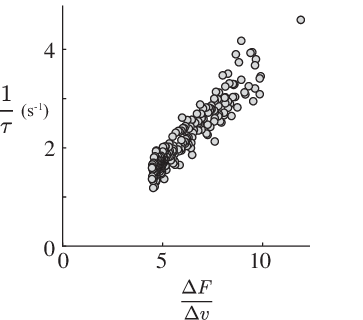}
    \caption{
        Experiment to measure the time to reach steady-state ($\tau)$ with differing approximately linear force-speed curves with effective viscosity of $\frac{\Delta F}{\Delta V}$ (See Fig.~\ref{fig:force_vel}b). 
    Plot shows the inverse of the startup time versus the effective viscosity parameter for an active contact system.
     }
    \label{fig:control_viscous}
\end{figure}

\textit{Contact ``viscosity'' control--}
In experiment we set the range of active contacts speeds to be uniformly distributed over a desired range to achieve a  stepped linear force-speed step profile (Fig.~\ref{fig:control_viscous}). 
The effective viscosity of this force profile is calculated as $\frac{\Delta F}{\Delta v}$.
In 490 experiments we randomly varied the controlled force-speed slope and measured the system speed versus time while started from rest and allowed to reach steady-state. 

If the contact forces were governed by real viscous forces the system speed would exponentially approach steady-state with a time-scale $\tau_{visc} = \frac{m}{\Delta F / \Delta v}$ where $m$ is the system inertia (Fig.~\ref{fig:motivation}).
To compare our commanded ``viscosity'' with the measurement we fit an exponential curve to the speed versus time curve of the system and determined the steady-state time constant $\tau$.
The time-constant should scale as $\frac{1}{\tau} \propto \frac{\Delta F}{\Delta v}$ in a viscous system.
The experimental observations agreed well with the effective-viscosity prediction indicating that by shaping the force-speed curve through active contacts we can mimic viscous force in dry-friction sliding (Fig.~\ref{fig:control_viscous}).   
 
\textit{Sliding friction coefficient control--}
In a final set of experiments we demonstrate how the coefficient of sliding friction can be controlled by modulating active contacts.
In ground-reaction force measurements we demonstrated that by setting half the active contacts to $+v_{max}$ and half to $-v_{max}$ the ground reaction force is small and near zero.
This observation motivates a method to control the sliding friction of an active contact system. 
By setting pairs of active contacts to be at opposite speeds, that contact pair will generate net zero sliding force when the system speed lies within the range $-v_{max} < v < +v_{max}$.
Thus, in a system with $N$ total active contacts there can be up to $N/2$ contacts ``deactivated''.
Since the normal force is unchanged when the active contacts speeds are change, this method effectively changes the sliding friction coefficient, $\mu$.  

\begin{figure}[t]
    \centering
    \includegraphics[width=1\linewidth]{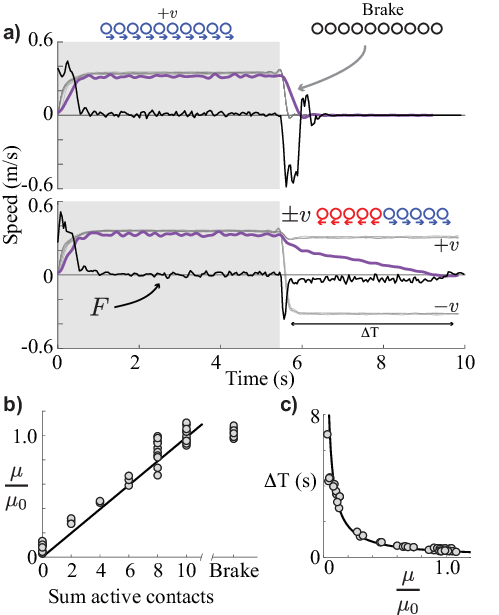}
    \caption{Demonstration of friction control from active contacts. 
    a) Active contact (gray lines) and system (purple line) speeds and friction force (black line) versus time for two experiments.
    The upper plot shows the case where active contacts are turned off, called the brake condition, and the system comes to rest from steady-state speed. 
    Note the associated large negative friction force in opposition to the positive direction of motion when the system is braked at the end of the gray shaded region.  
    The bottom plot shows the case where half active contacts are set to $-v_{max}$ and half at $+v_{max}$. 
    At the initiation of the active contact speed change (at the end of the gray shaded region) there is an initial force transient followed by a small friction force as the system gradually slides to rest.
    b) The friction coefficient as a function of the sum of positive and negative speed active contacts. 
    Coefficient is normalized to the value measured when the system is braked by setting all active contacts to $v = 0$.
    c) The transition time $\Delta T$ between steady-state speeds versus the measured effective friction coefficient across all experiments. 
    Black curve is the function $\Delta T = \frac{0.35}{\mu / \mu_0}$. 
     }
    \label{fig:coefficient}
\end{figure}

To test this hypothesis in experiment we first set all the active contact speeds to be equal and positive at $v_i = +v_{max}$ and let the system speed increase from rest to the steady-state $V = +v_{max}$. 
Once at steady-state, we next commanded $5 - n$ of the active contacts to $+v_{max}$ and $5 + n$ of the contacts to $-v_{max}$.
The rationale for this is that by varying $n \in [0, 1, 2, 3, 4, 5]$ should produce a controlled sliding friction coefficient ranging from $\mu = \mu_0 [0, 0.2, 0.4, 0.6, 0.8, 1]$ respectively (Figure~\ref{fig:coefficient}).

Changing the speed of the active contacts causes the system to approach a new steady-state speed while sliding. 
We measured the ground reaction force as the system transitions to the new steady state speed.
We then calculate the change in sliding friction coefficient as the ratio of $\mu / \mu_0 = F/F_0$ where $F_0$ is the friction force of the sliding system when the active contacts are all set to $v_i = 0$ (the brake condition in Fig.~\ref{fig:coefficient}). 
We observed good agreement between the model prediction and the measured ground reaction forces. 
At the extreme of setting equal forward and reverse contacts ($n=0$) we measured a net reduction of the coefficient of friction by $91\pm 4$\%. 

Similarly, we measured the time for the system to transition between the steady-state speeds ($\Delta T$). 
For a constant friction force causing a change in speed from $+v_{max}$ to $-v_{max}$ the time-scale for this should vary inversely proportional to friction coefficient, $\Delta T \propto \frac{1}{\mu/\mu_0}$. 
We similarly observe good agreement between this prediction and our measured steady-state transition time (Fig.~\ref{fig:coefficient}). 
Both experimental observations indicate that by controlling active contact pairs we can selectively modify the sliding friction of the system in real-time. 

\textit{Discussion and Conclusion--}
In this work we have studied the sliding friction of a body that can independently control the speed of its contact points, called an ``active contact'' system \cite{Chong2023-gy}.
The physics of active contacts are important in understanding frictional locomotion systems, which encompass nearly all terrestrial animals and ground-based robots. 
The fundamental novelty of active contact systems are the ability to generate controllable force-speed friction behavior from individual speed-independent forces. 
Prior work has illustrated how periodic motion of active contacts can generate effective viscous-like forces \cite{Chong2023-la} through time-averaging of the contacts over a gait. 
In this manuscript we demonstrated that multiple active contacts at different speeds can create an ``instantaneous'' force-speed curve without the need for averaging. 
Thus, we predict that in terrestrial locomotion systems with large friction compared to inertial forces (called the coasting number \cite{Rieser2024-yz}), multi-contact gaits will establish an instantaneous force-speed relationship that may change throughout the gait.
Beyond just the effective viscous behavior of active contact systems our experiments have demonstrated that a much broader repertoire of force-speed behavior can be achieved through the control of active contacts.

It is instructive to consider how the physics of active contact frictional systems can be described by the set properties of active contact speeds. 
For example, we have demonstrated that the equilibrium speed of an active contact system with identical friction forces at each contact is determined by the median of the set of active contact speeds (Fig.~\ref{fig:equilib}).  
We now propose that the force-speed curves of an active contact system can be described through the statistical distribution of the active contact speed set. 
Formally we posit that the force-speed curve of an active contact system with identical friction forces is related to the cumulative distribution function ($\text{CDF}(V)$) of the active contact speeds through the following relationship 
\begin{align}
    F(V) = \mu m g (1 - 2 \text{CDF}(V)) 
    \label{eq:CDF}
\end{align}
We directly demonstrate this in the case of a discrete set of $N$ active contacts speeds, $v_i$, by expressing the discrete probability density of contact speeds as $P(V) = \frac{1}{N}\sum_{i=1}^{N}\delta(V - v_i)$ where $\delta(x)$ is the Dirac delta function.
Integrating the probabiltiy distribution of the contact speeds yields the cumulative distribution function $\text{CDF}(V) = \int_{-\infty}^{\infty} P(V) = \frac{1}{N} \sum_{i=1}^{N} \frac{1}{2}\left(1 + \frac{V - v_i}{|V - v_i|}\right)$. 
Through subsitution it is easy to show that the standard equation for friction (Eq.~\ref{eq:friction}) is equal to our probabilistic definition (Eq.~\ref{eq:CDF}).  
This re-expression of the force-speed relationship is more than just a convenient notational change, it enables an analytical method for directly calculating $F(V)$ in continuous active contact systems such as rotating and sliding disks \cite{Farkas2003-hc} or slithering snakes \cite{Hu2009-kq} moving over a rigid surface. 
By extension, the drag forces of slow movement within granular material (GM) are speed independent \cite{Albert1999-ao} and occur over the continuous surface of the body, and thus the motion of active bodies moving within GM \cite{Maladen2009-uq, Rieser2024-yz, Gart2020-ap, Li2013-ap} may possibly be described by the same physics of continuous active contact systems. 

While our model and experiment are in good agreement, illustrating the validity of this thought-experiment approach, we note that the experiment is substantially simplified from the real-world cases of active contacts such as locomotion. 
For example, system in the real-world may have active contacts with time varying coefficients of friction or applied normal forces.
In this case we now have to consider each active contact, $v_i$, with its own associated friction force, $f_i$. 
By treating $f_i$ as weights on the probability distribution of contact speeds we can still solve for equilibrium speed as the weighted median of the set, and the force-speed curve by replacing the $\text{CDF}(V)$ in Equation~\ref{eq:CDF} with the weighted cumulative distribution function. 
Thus, while our demonstrated experiments are on a relatively simple platform the ideas can be translated to more complex active contact systems through the properties of weighted sets. 

The dependence of friction on non-uniform contact speeds has been observed in prior experiments. 
For example, a disk-sliding on a surface will slide over a longer distance (and thus experience less frictional force) if it is also spinning about the vertical axis \cite{Farkas2003-hc}. 
Similarly, a wine cork is easier to move along the axial direction when the cork is rotated: rotation redirects the frictional force along the tangential direction and thus reduces the axial force component, a property that has been used to allow robots to more easily insert pegs into holes \cite{Liu2019-uv}. 
In the experiments of this paper we have demonstrated how the control of active contact speeds can be used to modulate, in real-time, the frictional interactions between sliding objects.
It is estimated that 30\% of the worlds energy is lost to sliding friction per year \cite{Liu2022-dt}, thus understanding and controlling sliding friction has extreme importance.  
Beyond just an oddity of locomotion or frictional mechanics, this work provides new understanding of the physics of active contact systems and provides new inspiration for engineered solutions for friction control.

\textit{Acknowledgements--}
We thank Baxi Chong, Dan Goldman, Glenna Clifton, Brian Bittner, Dan Zhao, and Shai Revzen for helpful conversations.
We acknowledge funding from NSF-CAREER \#2048235. 

\bibliography{refs.bib}
 
\end{document}